\documentstyle[preprint,prb,aps]{revtex}

\begin{document}

\begin{titlepage}

\title
{Phase interference in
antiferromagnetic quantum tunneling with an arbitrarily directed magnetic field}

\author{Rong L\"{u}\footnote {Author to whom 
the correspondence should be addressed.\\
Electronic address: rlu@castu.tsinghua.edu.cn}, Hui Pan, Jia-Lin Zhu, 
and Bing-Lin Gu} 
\address{Center for Advanced Study, 
Tsinghua University, Beijing 100084, People's Republic of China
}
\date{\today}
\maketitle
\begin{abstract}
The quantum interference effects induced by the topological
phase are studied analytically
in biaxial antiferromagnets
with an external
magnetic field at an arbitrarily angle.
This study provides a nontrivial generalization of the Kramers degeneracy
for equivalent double-well system
to coherently spin tunneling at ground states as well as low-lying excited states
for antiferromagnetic system with asymmetric twin barriers.
The spin-phase interference effects are found to depend
on the orientation of the magnetic field distinctly.

\noindent
{\bf PACS number(s)}:  \\
03.65.Vf Phase: geometric; dynamic or topological\\
75.45.+j Macroscopic quantum phenomena in magnetic systems\\
75.50.Ee Antiferromagnetics\\
73.40.Gk Tunneling
\end{abstract}

\end{titlepage}

One recent experiment on the molecular magnets Fe$_8$ showed a direct
evidence of the topological part of the quantum spin phase (Berry phase) in
the spin dynamics.\cite{1} The importance of the topological Berry phase in
spin tunneling was elucidated by Loss {\it et al., }and von Delft and Henly.%
\cite{2} They showed that this term can lead to destructive (for
half-integer total spins) and instructive (for integer spins) interference
between opposite winding tunneling paths in single-domain ferromagnetic (FM)
particles. While spin-parity effects are sometimes be related to Kramers
degeneracy,\cite{2} they typically go beyond the Kramers theorem in a rather
unexpected way.\cite{3} Similar spin-parity effects were found in
antiferromagnetic (AFM) particles, where only the integer excess spins can
tunnel but not the half-integer ones.\cite{4} The effects of magnetic field
along the hard\cite{5} and medium\cite{6} axis were studied in AFM particles.

Theoretical results\cite{4,5,6} showed that in AFM particles the exchange
energy is enhanced to the magnetic anisotropy, which leads to tunneling of
the N\'{e}el vector a much stronger effect than tunneling of magnetization
in FM particles. Therefore, the AFM particle is expected to be a better
candidate for observing quantum tunneling than the FM particle with a
similar size. Up to now theoretical studies on AFM tunneling\cite{4,5,6}
have been focused on spin-phase interference between two opposite
ground-state tunneling paths. The spin-phase interference between
excited-level tunneling paths are unknown for AFM particles. Moreover, the
previous works\cite{4,5,6} have been confined to the condition that the
magnetic field be applied along the easy, medium, or hard axis, separately.
The purpose of this letter is to study the resonant quantum tunneling and
spin-phase interference at excited levels in AFM particles placed in a
magnetic field at an arbitrary angle $\theta _H$. Therefore,{\it \ our study
provides a nontrivial generalization of the Kramers degeneracy for
equivalent double-well system to coherently spin tunneling at ground states
as well as low-lying excited states for AFM\ system with asymmetric twin
barriers caused by the arbitrarily directed magnetic field.}

The system of interest is a small ($\sim 5$nm in\ radius), single-domain,
AFM particle at a temperature well below its anisotropy gap. According to
the two-sublattice model,\cite{4} there is a strong exchange energy ${\bf m}%
_1\cdot {\bf m}_2/\chi _{\bot }$ between the two sublattices, where ${\bf m}%
_1$ and ${\bf m}_2$ are the magnetization vectors of the two sublattices
with large, fixed and unequal magnitudes. In the following, we assume that $%
m_1>m_2$ and $m=m_1-m_2\ll m_1$. The system has the biaxial symmetry, with $%
\widehat{x}$ being the easy axis, $\widehat{y}$ being the medium axis, and $%
\widehat{z}$ being the hard axis. The magnetic field is applied in the $ZY$
plane, at an arbitrary angle in the range of $0\leq \theta _H<\pi /2$. In
order to obtain the tunnel splitting for quantum coherence, we shall
calculate the path integral: $\int {\cal D}\{\theta \}{\cal D}\{\phi \}\exp
[-{\cal S}_E(\theta ,\phi )]$, where ${\cal S}_E$ is the effective Euclidean
action for AFM tunneling, 
\begin{eqnarray}
{\cal S}_E(\theta ,\phi ) &=&\frac V\hbar \int d\tau \left\{ i\frac{m_1+m_2}%
\gamma \left( \frac{d\phi }{d\tau }\right) -i\frac m\gamma \left( \frac{%
d\phi }{d\tau }\right) \cos \theta +\frac{\chi _{\bot }}{2\gamma ^2}\left( 
\frac{d\theta }{d\tau }\right) ^2\right.  \nonumber \\
&&\left. +\frac{\chi _{\bot }}{2\gamma ^2}\left( \frac{d\phi }{d\tau }%
-i\gamma H_z\right) ^2\sin ^2\theta +E\left( \theta ,\phi \right) \right\} .
\eqnum{1}
\end{eqnarray}
$E\left( \theta ,\phi \right) =E_a\left( \theta ,\phi \right) -mH_z\cos
\theta -mH_y\sin \theta \sin \phi $, and $E_a\left( \theta ,\phi \right) $
is the magnetocrystalline anisotropy energy. $V$ is the volume of the AFM
particle, and $\gamma $ is the gyromagnetic ratio. $\theta $ and $\phi $ are
the angular components of ${\bf m}_1$, which can also determine the
direction of the N\'{e}el vector. $\tau =it$, and $m=\hbar \gamma s/V$,
where $s$ is the excess spin of the AFM particle due to the non-compensation
of two sublattices. For this case, the magnetocrystalline anisotropy energy
is $E_a\left( \theta ,\phi \right) =K_{\bot }\cos ^2\theta +K_{\Vert }\sin
^2\theta \sin ^2\phi $, where $K_{\Vert }$ and $K_{\bot }$ are the
longitudinal and the transverse anisotropy coefficients satisfying $K_{\bot
}\gg K_{\Vert }>0$. Therefore, the N\'{e}el vector is forced to lie in the $%
\theta =\theta _0$ plane, and the fluctuations of $\theta $ about $\theta _0$
are small. Introducing $\theta =\theta _0+\alpha $, $\left| \alpha \right|
\ll 1$, $E\left( \theta ,\phi \right) $ reduces to 
\begin{eqnarray}
E\left( \alpha ,\phi \right) &=&K_{\bot }\sin ^2\theta _0\alpha ^2+K_{\Vert
}\sin ^2\theta _0\left( \sin \phi -\sin \phi _0\right) ^2  \nonumber \\
&&+2K_2\sin \theta _0\cos \theta _0\left( \sin \phi -\sin \phi _0\right)
^2\alpha ,  \eqnum{2}
\end{eqnarray}
where $\cos \theta _0=mH_z/2K_{\bot }$, $\sin \phi _0=mH_y/2K_{\Vert }\sin
\theta _0=h\sin \theta _H/\sqrt{1-\left( \lambda h\cos \theta _H\right) ^2}$%
, $\lambda =K_{\Vert }/K_{\bot }$, $h=H/H_0$, and $H_0=2K_{\Vert }/m$.

In the semiclassical limit, the dominant contribution to the transition
amplitude comes from finite action solution (instanton) of the classical
equation of motion. Since the configuration space of this problem is a
circle, only two types of instantons must be taken into account. We use $A$
to denote the instanton passing through the barrier at $\theta =\theta _0$, $%
\phi =\pi /2$, and $B$ as the instanton passing through the barrier at $%
\theta =\theta _0$, $\phi =3\pi /2$. Correspondingly, there are two kinds of
anti-instantons: $A^{-}$ and $B^{-}$. The small barrier at $\phi =\pi /2$ is 
$\hbar U_S=\hbar U\left( \phi =\pi /2\right) =K_{\Vert }V\sin ^2\theta
_0\left( 1-\sin \phi _0\right) ^2$, and the large barrier at $\phi =3\pi /2$
is $\hbar U_L=\hbar U\left( \phi =3\pi /2\right) =K_{\Vert }V\sin ^2\theta
_0\left( 1+\sin \phi _0\right) ^2$. Performing the Gaussian integration over 
$\alpha $, we can map the spin system onto a particle moving problem in
one-dimensional potential well. Now the Euclidean transition amplitude of
this system becomes 
\begin{eqnarray}
{\cal K}_E &=&\exp \left\{ -i\left[ S_{tot}-\left( 1+\frac \lambda 2+\frac %
\lambda 2\sin ^2\phi _0\right) s\cos \theta _0-2S\left( \frac Ss\right)
\left( \frac{K_{\bot }}J\right) \cos \theta _0\sin ^2\theta _0\right] \left(
\phi _f-\phi _i\right) \right\}  \nonumber \\
&&\times \int d\phi \exp \left\{ -\int d\tau \left[ \frac 12{\cal M}\left( 
\frac{d\phi }{d\tau }\right) ^2+U\left( \phi \right) \right] \right\} , 
\eqnum{3}
\end{eqnarray}
where $S_{tot}=2S-s$ is the total spins of two sublattices, and $%
S=m_1V/\hbar \gamma $ is the sublattice spins. In Eq. (3) we have taken $%
\chi _{\bot }=m_1^2/J$, where $J$ is the exchange interaction between two
sublattices. The effective mass and effective potential in Eq. (3) are 
\begin{eqnarray}
{\cal M} &=&\frac{\hbar S^2}{JV}\sin ^2\theta _0\left[ 1+\frac 1{2\sin
^2\theta _0}\left( \frac J{K_{\bot }}\right) \left( \frac sS\right)
^2\right] ,  \eqnum{4a} \\
U\left( \phi \right) &=&2\frac{K_{\Vert }V}\hbar \sin ^2\theta _0\left( \sin
\phi -\sin \phi _0\right) ^2.  \eqnum{4b}
\end{eqnarray}
The potential $U\left( \phi \right) =U\left( \phi +2n\pi \right) $ has an
asymmetric twin barrier. The degenerate ground states are given by two
different types of minima of $U\left( \phi \right) $ at $2n\pi +\phi _0$ and 
$\left( 2n+1\right) \pi -\phi _0$. $U\left( \phi \right) $ can be regarded
as a superlattice consisting of two sublattices, and the energy spectrum can
be obtained by applying the Bloch theorem and the tight-binding
approximation. The translational symmetry is ensured by the possibility of
successive $2\pi $ extensions.

The periodic instanton configuration $\phi _p$ satisfies the equation of
motion $\frac 12{\cal M}\left( \frac{d\phi _p}{d\tau }\right) ^2-U\left(
\phi _p\right) =-E$, where $E>0$ can be viewed as the classical energy of
the pseudoparticle configuration. Then the periodic instanton $A$ solution
for $0\leq E\leq U_S$ is $\sin \phi _A=\frac{1-\xi _1\text{sn}^2\left(
\omega _1\tau ,k_1\right) }{1+\xi _1\text{sn}^2\left( \omega _1\tau
,k_1\right) }$. sn$\left( \omega _1\tau ,k_1\right) $ is the Jacobian
elliptic sine function of modulus $k_1=\sqrt{\frac{\left( 1-\alpha \right)
\left( 1+\beta \right) }{\left( 1+\alpha \right) \left( 1-\beta \right) }}$,
where $\alpha =\sin \phi _0+\sqrt{\frac{\hbar E}{K_{\Vert }V\sin ^2\theta _0}%
}$, $\beta =\sin \phi _0-\sqrt{\frac{\hbar E}{K_{\Vert }V\sin ^2\theta _0}}$%
, $\xi _1=\left( 1-\alpha \right) /\left( 1+\alpha \right) $, $\omega
_1=\omega _0/g_1$, $\omega _0=\sqrt{2K_{\Vert }V/\hbar {\cal M}}\sin \theta
_0$, and $g_1=2/\sqrt{\left( 1+\alpha \right) \left( 1-\beta \right) }$. The
associated Euclidean action is 
\begin{equation}
{\cal S}_A=\int_{-\beta }^\beta d\tau \left[ \frac 12{\cal M}\left( \frac{%
d\phi _A}{d\tau }\right) ^2+U\left( \phi _A\right) \right] ={\cal W}%
_A+2E\beta ,  \eqnum{5a}
\end{equation}
where 
\begin{equation}
{\cal W}_A=4{\cal M}\omega _1\left[ E\left( k_1\right) +\frac{\left(
k_1^2-\xi _1\right) }{\xi _1}K\left( k_1\right) +\frac{\left( \xi
_1^2-k_1^2\right) }{\xi _1}\Pi \left( k_1,\xi _1\right) \right] .  \eqnum{5b}
\end{equation}
Here $K\left( k_1\right) $, $E\left( k_1\right) $, and $\Pi \left( k_1,\xi
_1\right) $ are the complete elliptic integral of the first, second, and
third kind, respectively. The similar method can be applied to the periodic
instanton $B$, and the result is ${\cal S}_B={\cal W}_B+2E\beta $ for $0\leq
E\leq U_S$, where 
\begin{equation}
{\cal W}_A=4{\cal M}\omega _1\left[ E\left( k_1\right) +\frac{\left(
k_1^2-\xi _1\right) }{\xi _1}K\left( k_1\right) +\frac{\left( \xi
_1^2-k_1^2\right) }{\xi _1}\Pi \left( k_1,\xi _1\right) \right] ,  \eqnum{6}
\end{equation}
and $\xi _2=\left( 1+\beta \right) /\left( 1-\beta \right) $. While for $%
U_S\leq E\leq U_L$, the result is $\widetilde{{\cal S}}_B=\widetilde{{\cal W}%
}_B+2E\beta $, where 
\begin{equation}
\widetilde{{\cal W}}_B=2{\cal M}\xi _3\left( 1+\alpha \right) \omega
_2\left[ \frac 1{k_2^2-\xi _3}E\left( k_2\right) -\frac 1{\xi _3}K\left(
k_2\right) +\frac{k_2^2+\xi _3^2+2k_2^2\xi _3}{\xi _3}\Pi \left( k_2,\xi
_3\right) \right] ,  \eqnum{7}
\end{equation}
with $\omega _2=\omega _0/g_2$, $\xi _3=\left( 1+\beta \right) /\left(
\alpha -\beta \right) $, $g_2=\sqrt{2/\left( \alpha -\beta \right) }$, and $%
k_2^2=\frac{\left( \alpha -1\right) \left( 1+\beta \right) }{2\left( \alpha
-\beta \right) }$.

Now we turn to the calculation of level splittings of excited states. For a
particle moving in a double-well-like potential $U\left( x\right) $, the WKB
formula gives the tunnel splittings of the $n$th degenerate excited levels
or the imaginary parts of the $n$th metastable levels as $\Delta E_n\left( 
\text{or }%
\mathop{\rm Im}%
E_n\right) =\frac{\omega \left( E_n\right) }\pi \exp \left( -{\cal W}\right) 
$,\cite{7,8} where $\omega \left( E_n\right) =2\pi /t\left( E_n\right) $. $%
t\left( E_n\right) =\sqrt{2m}\int_{x_1\left( E_n\right) }^{x_2\left(
E_n\right) }\frac{dx}{\sqrt{E_n-U\left( x\right) }}$ is the period of the
real-time oscillation in the potential well, where $x_{1,2}\left( E_n\right) 
$ are the turning points for the particle oscillating inside the potential $%
U\left( x\right) $. For the present case, we find that the level splittings
for instantons $A$ and $B$ in the domain $0\leq E\leq U_S$ are $\Delta {\cal %
E}_{A\left( B\right) }=\frac 2{t_{A\left( B\right) }\left( E\right) }\exp
\left( -{\cal W}_{A\left( B\right) }\right) $, where $t_A\left( E\right)
=t_B\left( E\right) =\frac 2{\omega _1\left( E\right) }K\left( k_1^{\prime
}\right) $, and $k_1^{\prime }=\sqrt{1-k_1^2}$. For $U_S\leq E\leq U_L$, the
imaginary parts of the metastable energy levels are $%
\mathop{\rm Im}%
E=\frac 2{\widetilde{t}_B\left( E\right) }\exp \left( -2\widetilde{{\cal W}}%
_B\right) $, where $\widetilde{t}_B\left( E\right) =\frac 2{\omega _2\left(
E\right) }K\left( k_2^{\prime }\right) $, and $k_2^{\prime }=\sqrt{1-k_2^2}$.

Then we discuss the low energy limit of the level splitting. By using the
small oscillator approximation for energy near the bottom of the potential
well, ${\cal E}_n=\left( n+1/2\right) \Omega $, $\Omega =\sqrt{\left(
d^2U/d\phi ^2\right) _{\phi =\phi _0}/{\cal M}}=\sqrt{2K_{\Vert }V/\hbar 
{\cal M}}\sin \theta _0\cos \phi _0$, Eqs. (5b) and (6) are expanded as 
\begin{eqnarray}
{\cal W}_{A\left( B\right) ,n} &=&{\cal W}_{A\left( B\right) ,0}-\left( n+%
\frac 12\right) +\left( n+\frac 12\right)  \nonumber \\
&&\times \ln \left( \frac{n+1/2}{2^{7/2}\sqrt{\frac{K_{\Vert }}J}S\sqrt{1+%
\frac 1{2\sin ^2\theta _0}\left( \frac J{K_{\bot }}\right) \left( \frac sS%
\right) ^2}\sin ^2\theta _0\cos ^3\phi _0}\right) ,  \eqnum{8a} \\
{\cal W}_{A\left( B\right) ,0} &=&2^{3/2}\sqrt{\frac{K_{\Vert }}J}S\sin
^2\theta _0\sqrt{1+\frac 1{2\sin ^2\theta _0}\left( \frac J{K_{\bot }}%
\right) \left( \frac sS\right) ^2}  \nonumber \\
&&\times \left( \cos \phi _0\mp 2\sin \phi _0\arcsin \sqrt{\frac{1-\sin \phi
_0}2}\right) ,  \eqnum{8b}
\end{eqnarray}
where ``$-$'' for the instanton $A$, and ``+'' for the instanton $B$.
Therefore, the tunnel splittings of $n$th excited levels are found to be 
\begin{eqnarray}
\hbar \Delta {\cal E}_{A\left( B\right) ,n} &=&\frac{2\cos \phi _0}{\sqrt{%
\pi }n!}\frac{\sqrt{K_{\bot }J}V}S\frac 1{\sqrt{1+\frac 1{2\sin ^2\theta _0}%
\left( \frac J{K_{\bot }}\right) \left( \frac sS\right) ^2}}\left[ 2^{7/2}%
\sqrt{\frac{K_{\Vert }}J}S\right.  \nonumber \\
&&\left. \times \sqrt{1+\frac 1{2\sin ^2\theta _0}\left( \frac J{K_{\bot }}%
\right) \left( \frac sS\right) ^2}\cos ^3\phi _0\sin ^2\theta _0\right]
^{n+1/2}\exp \left( -{\cal W}_{A\left( B\right) ,0}\right) .  \eqnum{9}
\end{eqnarray}
Eq. (8b) shows that at finite magnetic field the WKB exponent for instanton $%
A$, ${\cal W}_{A,0}$ is smaller than that for instanton $B$, ${\cal W}_{B,0}$
because the barrier through which instanton $B$ must tunnel is higher than
that for instanton $A$.

It is noted that $\hbar \Delta {\cal E}_{A,n}$ or $\hbar \Delta {\cal E}%
_{B,n}$ is only the level shift induced by tunneling between degenerate
excited states through a single barrier. The periodic potential $U\left(
\phi \right) $ can be regarded as a one-dimensional superlattice with the
sublattices $A$ and $B$. The general translation symmetry results in the
energy ``band'' structure which is formally the same as that of a
one-dimensional tight-binding model in solid state physics, and the energy
spectrum could be determined by the Bloch theorem. The Bloch states for
sublattices $A$ and $B$ can be written as $\Phi _A\left( \xi ,\phi \right) =%
\frac 1{\sqrt{L}}\sum_ne^{i\xi \phi _n}\varphi _A\left( \phi -\phi _n\right) 
$, and $\Phi _B\left( \xi ,\phi \right) =\frac 1{\sqrt{L}}\sum_ne^{i\xi
\left( \phi _n+a\right) }\varphi _B\left( \phi -\phi _n-a\right) $, where $%
\phi _n=2n\pi +\phi _0$, $L=N\left( a+b\right) $, $a=\pi -2\phi _0$, and $%
b=\pi +2\phi _0$. Then the total wavefunction $\Psi _\xi \left( \phi \right) 
$ is a linear combination of the two Bloch states, $\Psi _\xi \left( \phi
\right) =a_A\left( \xi \right) \Phi _A\left( \xi ,\phi \right) +a_B\left(
\xi \right) \Phi _B\left( \xi ,\phi \right) $. Including the phase
contributions of the topological term, we derive the secular equation as 
\begin{equation}
\left[ 
\begin{array}{ll}
{\cal E}_n-E\left( \xi \right) & e^{i\left( \xi -\mu \right) a}\Delta {\cal E%
}_{A,n}+e^{-i\left( \xi -\mu \right) b}\Delta {\cal E}_{B,n} \\ 
e^{-i\left( \xi -\mu \right) a}\Delta {\cal E}_{A,n}+e^{i\left( \xi -\mu
\right) b}\Delta {\cal E}_{B,n} & {\cal E}_n-E\left( \xi \right)
\end{array}
\right] \left[ 
\begin{array}{l}
a_A\left( \xi \right) \\ 
a_B\left( \xi \right)
\end{array}
\right] =0,  \eqnum{10a}
\end{equation}
where 
\begin{equation}
\mu =S_{tot}-\left( 1+\frac \lambda 2+\frac \lambda 2\sin ^2\phi _0\right)
s\cos \theta _0-2S\left( \frac Ss\right) \left( \frac{K_{\bot }}J\right)
\cos \theta _0\sin ^2\theta _0,  \eqnum{10b}
\end{equation}
and ${\cal E}_n=\left( n+1/2\right) \Omega $. The Bloch wave vector $\xi =0$
in the first Brillouin zone. Therefore, the eigenvalues of Eq. (10a) are 
\begin{equation}
E_{\pm }={\cal E}_n\pm \sqrt{\left( \Delta {\cal E}_{A,n}\right) ^2+\left(
\Delta {\cal E}_{B,n}\right) ^2+2\left( \Delta {\cal E}_{A,n}\right) \left(
\Delta {\cal E}_{B,n}\right) \cos \Theta },  \eqnum{11a}
\end{equation}
where 
\begin{equation}
\Theta =2\pi \left\{ s\left[ 1+\left( 1+\frac \lambda 2+\frac \lambda 2\sin
^2\phi _0\right) \cos \theta _0\right] +2S\left( \frac Ss\right) \left( 
\frac{K_{\bot }}J\right) \cos \theta _0\sin ^2\theta _0\right\} . 
\eqnum{11b}
\end{equation}
The tunnel splitting of $n$th excited level is $\Delta {\cal E}_n=2\sqrt{%
\left( \Delta {\cal E}_{A,n}\right) ^2+\left( \Delta {\cal E}_{B,n}\right)
^2+2\left( \Delta {\cal E}_{A,n}\right) \left( \Delta {\cal E}_{B,n}\right)
\cos \Theta }$. We can rederive this result by calculating the transition
amplitude, or by the effective Hamiltonian method.\cite{9}

At zero magnetic field, $\sin \phi _0=0$, $\cos \theta _0=0$, $\Delta {\cal E%
}_{A,n}=\Delta {\cal E}_{B,n}$, the tunnel splitting is suppressed to zero
for the half-integer excess spins, which is in good agreement with the
Kramers theorem. The presence of a magnetic field perpendicular to the plane
of rotation of magnetization yields an additional contribution to the
topological phase, resulting constructive and destructive interferences
alternatively for both integer and half-integer excess spins. Tunneling is
thus periodically suppressed. At finite magnetic field, the tunneling
spectrum of the degenerate $n$th excited levels depends on the parity of
excess spins $s$, 
\begin{eqnarray}
\Delta {\cal E}_n &=&2\left\{ \left( \Delta {\cal E}_{A,n}\right) ^2+\left(
\Delta {\cal E}_{B,n}\right) ^2\pm 2\left( \Delta {\cal E}_{A,n}\right)
\left( \Delta {\cal E}_{B,n}\right) \right.  \nonumber \\
&&\left. \times \cos \left[ 2\pi \left( \left( 1+\frac \lambda 2+\frac %
\lambda 2\sin ^2\phi _0\right) s\cos \theta _0+2S\left( \frac Ss\right)
\left( \frac{K_{\bot }}J\right) \cos \theta _0\sin ^2\theta _0\right)
\right] \right\} ^{1/2},  \eqnum{12}
\end{eqnarray}
where ``$+$'' for integer $s$, and ``$-$'' for half-integer $s$. The
spin-parity effect and the oscillation of the tunnel splitting with the
filed is shown in Fig. 1. Another important observation is that only the $%
\widehat{z}$ component of the magnetic field (i.e., along the hard axis) can
lead to the oscillation of the tunnel splitting for the highly anisotropic
case. As shown in Fig. 2, even a small misalignment of the field with the $%
\widehat{z}$ axis can completely destroy the oscillation effect, and the
oscillation is absent when the field is along the medium axis. For small $%
\theta _H$ the tunnel splitting oscillates with the field, whereas no
oscillation is shown up for large $\theta _H$. In the latter case, a much
stronger increase of tunnel splitting with the field is shown. This strong
dependence on the orientation of the field can be observed for ground-state
resonance as well as excited-state resonance. As a result, we conclude that%
{\it \ the spin-phase interference effects depend on the orientation of the
external magnetic field distinctly. This distinct angular dependence,
together with the oscillation of the tunnel splittings with the field, may
provide an independent experimental test for the spin-phase interference
effects in AFM particles}. In Fig. 3, we plot the tunnel splittings of the
ground-state level and the first excited level as a function of the magnetic
field for integer excess spins. It is clearly shown that the splitting is
enhanced by quantum tunneling at the excited levels. Detailed calculations
of the thermodynamic quantities of the system show that the specific heats
oscillate with the magnetic field and are strongly parity dependent of
excess spins at sufficiently low temperatures. Due to the topological nature
of the Berry phase, these spin-parity effects are independent of details
such as the magnitude of excess spins, the shape of the soliton and the
tunneling potential.

More recently, Wernsdorfer and Sessoli have measured the tunnel splittings
in the molecular Fe$_8$ clusters with the help of an array of micro-SQUIDs,
and have found a clear oscillation in the tunnel splittings.\cite{1} {\it %
Similar spin-phase interference effects observed in ferromagnetic Fe}$_8$%
{\it \ cluster are found theoretically in this letter for AFM particles},
which may bring a new insight to test the manifestation of quantum effects
at a macroscopic level and the influence of quantum phases on the tunneling
bahaviors of spin systems.

\section*{Acknowledgments}

R. L. and J. L. Z. would like to thank Professor W. Wernsdorfer and
Professor R. Sessoli for providing their paper (Ref. 1).

Fig. 1 The relative tunnel splitting of the first excited level ($n=1$) $%
\Delta \varepsilon _1/\Delta \varepsilon _{A,0}\left( H=0\right) $ as a
function of $H/H_0$ for integer $\left( s=10\right) $ and half-integer $%
\left( s=10.5\right) $ excess spins at $\theta _H=0^{\circ }$. Here $S=1000$%
, $K_1/J=0.002$, and $\lambda =K_2/K_1=0.02$.

Fig. 2 The relative tunnel splitting of the first excited level ($n=1$) $%
\Delta \varepsilon _1/\Delta \varepsilon _{A,0}\left( H=0\right) $ as a
function of $H/H_0$ for $\theta _H=0^{\circ }$, $1^{\circ }$, $3^{\circ }$, $%
5^{\circ }$, $10^{\circ }$, $20^{\circ }$ and $90^{\circ }$, respectively.
Here $s=10$, $S=1000$, $K_1/J=0.002$, and $\lambda =K_2/K_1=0.02$.

Fig. 3 The relative tunnel splitting $\Delta \varepsilon _n/\Delta
\varepsilon _{A,0}\left( H=0\right) $ of the ground-state level ($n=0$) and
the first excited level ($n=1$) as a function of $H/H_0$ for integer $\left(
s=10\right) $ excess spins at $\theta _H=0^{\circ }$. Here $S=1000$, $%
K_1/J=0.002$, and $\lambda =K_2/K_1=0.02$.

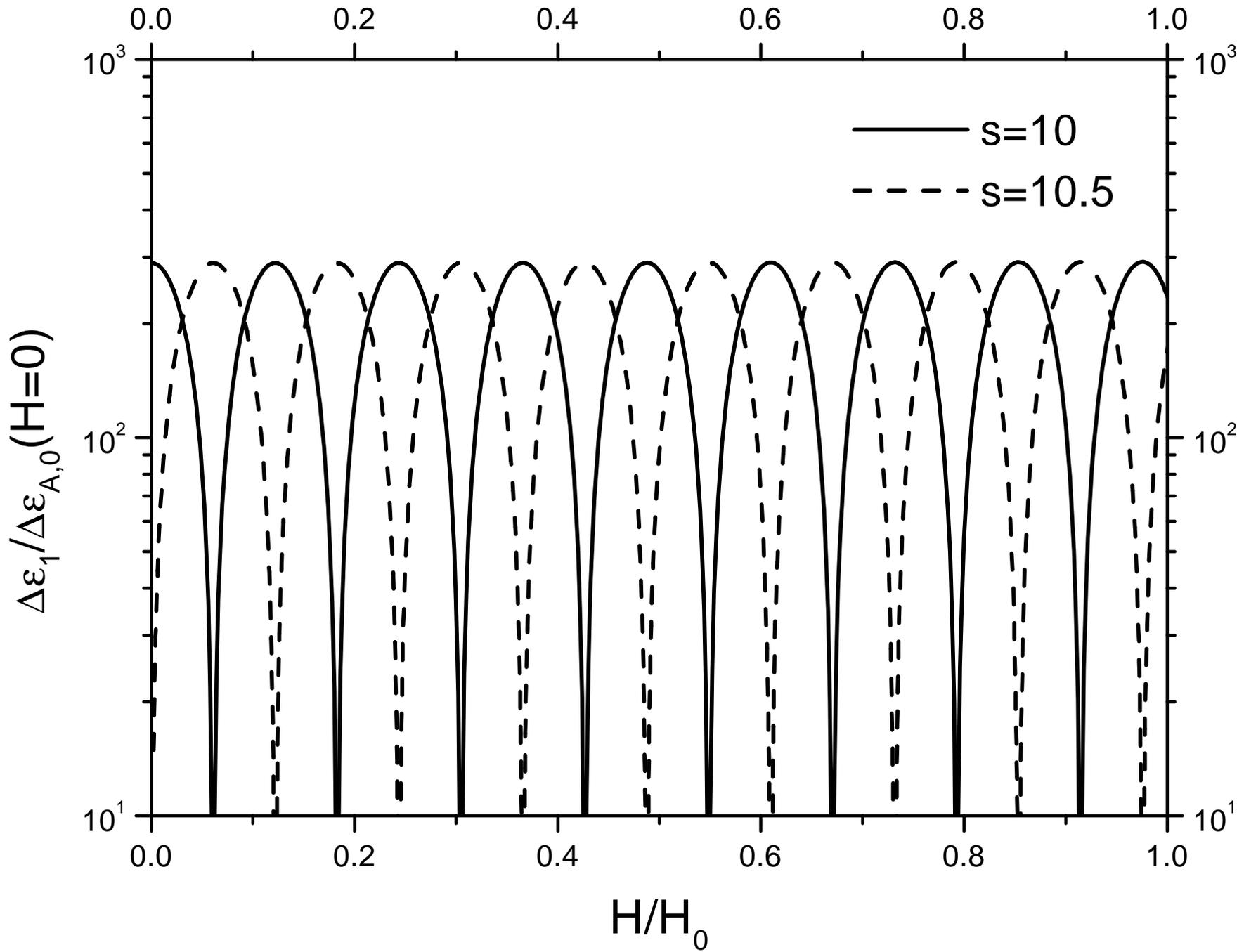
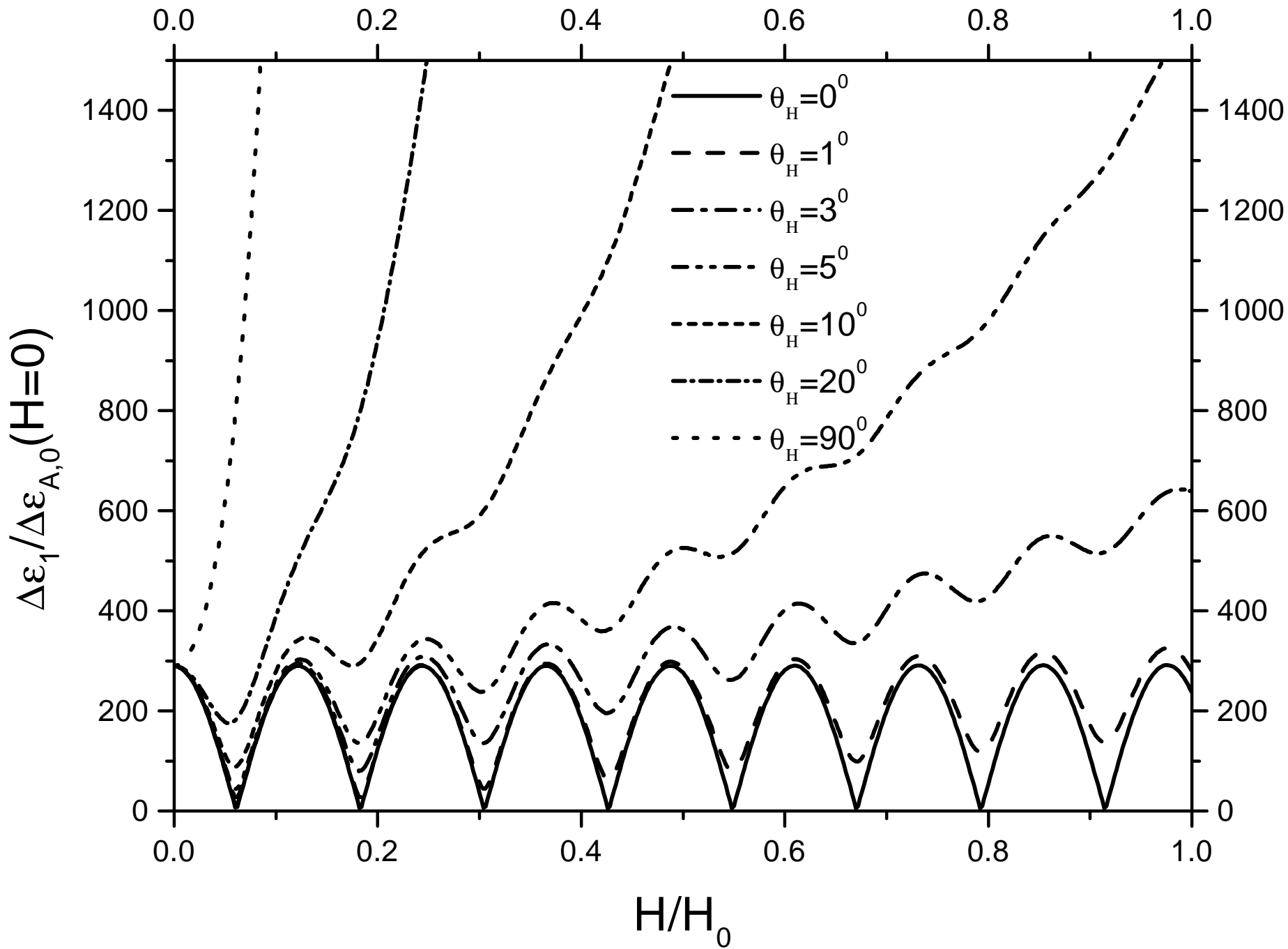
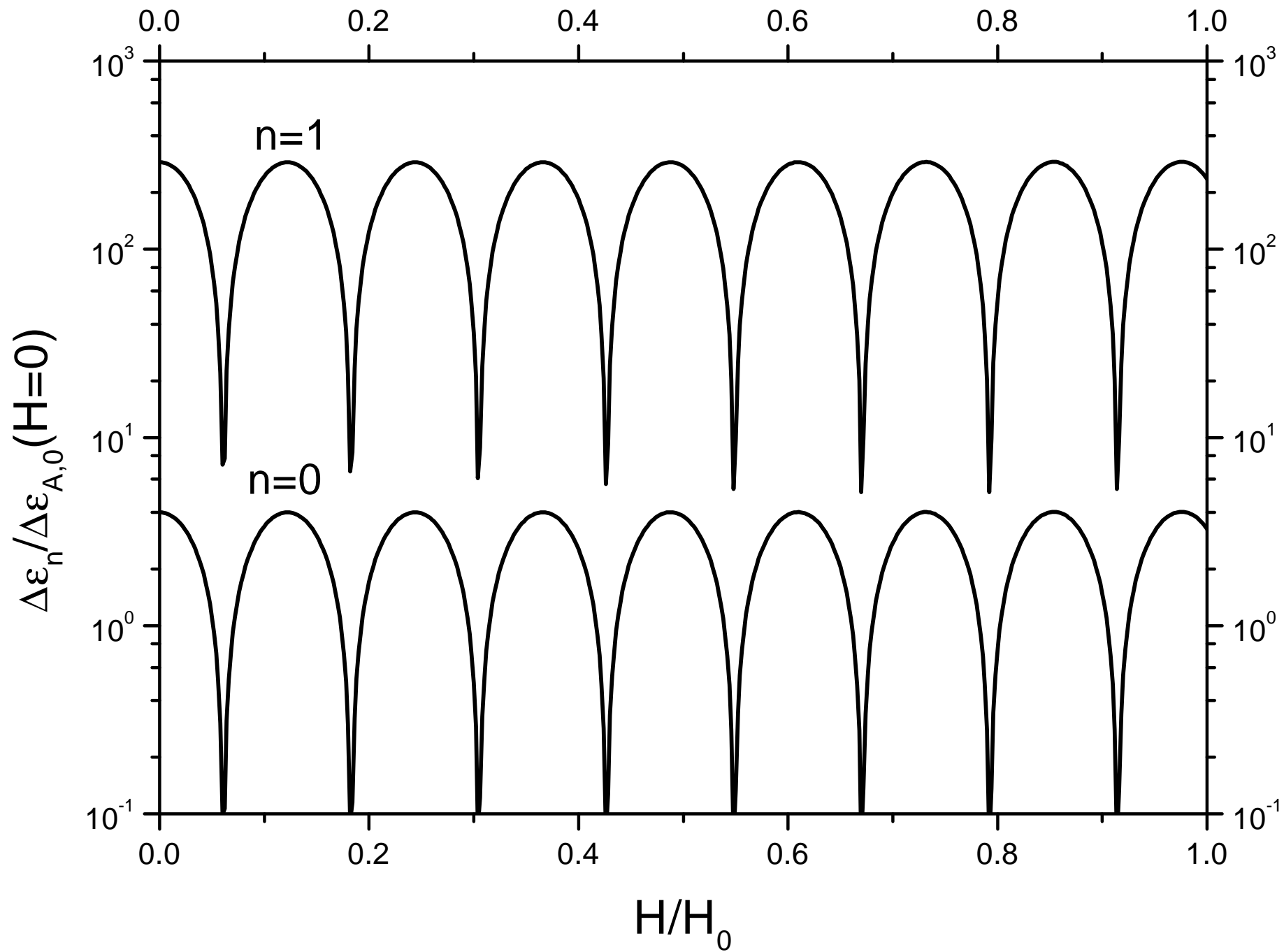
\end{document}